\documentclass[a4paper,11pt]{article}
\usepackage{verbatim}
\usepackage{textcomp}
\usepackage{amsmath}
\usepackage{mathrsfs}
\usepackage{amssymb}
\usepackage{amsthm}
\usepackage{slashed}
\usepackage[english]{babel}
\usepackage{graphicx}
\usepackage{makeidx}
\usepackage{appendix}
\usepackage[applemac]{inputenc}
\usepackage{tikz}
\usetikzlibrary{shapes,arrows}
\usepackage{listings}
\usepackage{sidecap}
\usepackage{wrapfig}
\newcommand{\bqa}{\begin{eqnarray}}
\newcommand{\eqa}{\end{eqnarray}}
\newcommand{\nl}{\nonumber \\}
\newcommand{\de}{\,\mathrm{d}}

\newcommand{\ep}{\,\epsilon}
\newcommand{\D}{\,{D}}
\newcommand{\Dbar}{\,\overline{{D}}}
\newcommand{\qbar}{\,\overline{q}}
\newcommand{\I}{\,\mathfrak{I}}

\newcommand{\amp}{\,\mathcal{M}}

\usetikzlibrary{decorations.pathreplacing}

\newcommand{\Mw}{M^2_W}

\newcommand{\FeynArts}{\emph{FeynArts }}

\newcommand{\be}{\,\begin{equation}}
\newcommand{\ee}{\,\end{equation}}
\newcommand{\bes}{\,\begin{equation*}}
\newcommand{\ees}{\,\end{equation*}}

\newcommand{\fdr}{\,[\de^4q]}
\newcommand{\Tr}{\,\text{Tr}}
\newcommand{\dr}{\,\frac{\de^nq}{\mu_R^{\ep}}}
\newcommand{\qbarslash}{\,\overline{\slashed{q}}}

\newcommand{\FF}{= i \frac{\slashed{p}+m_f}{p^2-m_f^2}}
\newcommand{\VFF}{=-ieQ_f\gamma^{\mu}}
\newcommand{\SFF}{= -i \frac{e \,m_f}{2s_WM_W}}

\newcommand{\VVVa}{= ie\,V_3^{\mu\rho\sigma}}

\newcommand{\VVVV}{= -ie^2\,V_4^{\mu\nu\rho\sigma} }

\newcommand{\ScSc}{=\frac{i}{p^2-\xi M_w^2}}

\newcommand{\VV}{=\frac{-i \Big[g^{\mu\nu} -(1-\xi) \frac{k^{\mu}k^{\nu}}{k^2-\xi M_W^2}\Big] }{k^2-M_W^2}}

\newcommand{\SVV}{=ieg^{\mu\nu}C_{SVV}}

\newcommand{\SSVV}{=ie^2g^{\mu\nu}C_{SSVV}}

\newcommand{\VSS}{=ie (p_1-p_2)^{\mu}\,C_{VS_1S_2}}

\newcommand{\SSS}{=-i\frac{e\,M_H^2}{2\,s_W\,M_W}}

\newcommand{\GG}{= \frac{i}{k^2-\xi M_W^2}}
\newcommand{\VGG}{=\pm i e p^{\mu} }
\newcommand{\SGG}{= -i \frac{e \,\xi M_W}{2s_W}}

\usepackage{axodraw4j}
\usepackage{pstricks}
\usepackage{color}

\usepackage{jheppub} 

\usepackage[T1]{fontenc} 
\title{Gauge invariance at work in FDR: \boldmath $H \to \gamma \gamma$}

\author{A. Donati and R. Pittau}
\affiliation{Departamento de F\'{\i}sica Te\'orica y del Cosmos,\\
Campus Fuentenueva s. n., Universidad de Granada 
E-18071 Granada, Spain}
\emailAdd{adonati@ugr.es}
\emailAdd{pittau@ugr.es}

\abstract{
We present the first complete calculation performed within the Four Dimensional Regularization scheme (FDR), namely the loop-induced on-shell amplitude for the Higgs boson decay into two photons in an arbitrary $R_\xi$ gauge. 
FDR is a new technique \mbox{-free} of infinities- for addressing multi-loop calculus, which automatically preserves gauge invariance, allowing for a 4-dimensional computation at the same time. 
We obtained the same result as that assessed in dimensional regularization, thereby explicitly verifying, in a realistic case, that FDR respects gauge invariance.
}

\begin{document} 

\newcommand{\mur}{\mu_{\scriptscriptstyle R}}

\maketitle
\flushbottom
\section{Introduction}
\label{sec:intro}
Calculating higher order corrections in Quantum Field Theories (QFTs) is 
becoming more and more influential both from a phenomenological point of view and from a more theoretical perspective. The success of the Standard Model (SM) in describing the observed high energy data, especially in the light of the discovery of a Higgs boson candidate~\cite{:2012gk,:2012gu}, leaves little room for large and unexpected signals at colliders, at least at the present energy regime. One is therefore forced to rely on the full predictive power of QFTs and
compute the tiny effects induced by the Radiative Corrections (RCs), in the hope of finding small deviations from the observed data.

 The computation of RCs is very demanding from a technical point of view. Methods used in the past decades for processes with moderate multiplicities are becoming less and less adequate in view of the growing complexity of the interesting final states. While a lot of work has been recently devoted to deal, in an efficient way, with 1-loop processes at large multiplicities~\cite{Ossola:2006us,Berger:2008sj,Giele:2008ve,Ellis:2011cr}, very little simplification has been achieved so far in the field of the multi-loop calculations \footnote{See, however,~\cite{Mastrolia:2011pr,Badger:2012dv,Johansson:2012zv,Kleiss:2012yv}.}.
What seems to trigger most of the difficulties is the usual treatment, in the framework of Dimensional Regularization (DR)~\cite{'tHooft:1972fi}, of the infinities arising in the intermediate steps of the calculation: DR forces a huge and cumbersome analytic work to be done to ensure all needed cancellations, even before starting the actual computation of the relevant physical (finite) contribution. Therefore it looks obvious that major simplifications are expected when performing the computations directly in the physical 4-dimensional Minkowsky space, also in the presence of Ultraviolet (UV) and Infrared/Collinear divergences. 
 This observation led to several attempts to find four-dimensional alternatives to the DR treatment of the UV infinities, such as differential renormalization~\cite{Freedman:1991tk}, constrained differential renormalization~\cite{delAguila:1997kw,delAguila:1998nd}, which both work in the coordinate space, implicit renormalization~\cite{Battistel:1998sz,Cherchiglia:2010yd} and LR~\cite{Wu:2003dd}, directly applicable in the momentum space.
 In a recent work, the FDR approach~\cite{Pittau:2012zd} has been proposed by one of us in which the UV problem is solved by simply re-interpreting the loop integrals appearing in the calculation. They are {\em defined} in such a way that infinities do not occur, at the price of introducing an arbitrary scale, called  $\mu$, which plays the role of the renormalization scale. The procedure works because the FDR re-interpretation respects gauge invariance by construction.  

 The purpose of this work is to illustrate some of the assets of the FDR scheme, and to show how to put it into practice. 
To this goal, we present the first application of FDR to a complete calculation in the EW theory. In order to keep the most general approach, we have decided to work in an arbitrary $R_{\xi}$-gauge, thereby explicitly verifying that the method respects gauge invariance.  We have chosen to compute the 1-loop on-shell amplitude for the Higgs boson decaying into two photons, which is  known since a long time~\cite{Ellis:1975ap,Ioffe:1976sd,Shifman:1979eb,Rizzo:1979mf} and, given its relevance and simplicity, has been recently reconsidered by several authors as a case study~\cite{Cherchiglia:2012zp,Shao:2011wx,Dedes:2012hf,Piccinini:2011az,Jegerlehner:2011jm,Huang:2011yf,Shifman:2011ri,Gastmans:2011wh,Gastmans:2011ks,Bursa:2011aa,Marciano:2011gm}. Because there is no $H\gamma\gamma$ interaction at tree level in the SM, the process is finite; although no renormalization is needed, infinities arise at intermediate steps of the calculation, which demands to work within a divergence-safe framework, such as FDR. Moreover, being indirect, the process is mediated by either bosonic or fermionic loops. 
Calculating the bosonic contribution stands as a strong test of the gauge invariance property of FDR. On the other hand, the fermionic contribution gives us the opportunity of illustrating FDR in the presence of fermionic loops. 

The outline of the paper is as follows: in section~\ref{The_FDR_method} we review the FDR method, in section~\ref{PV_reduction_in_FDR} we present tensor reduction in FDR and, in section~\ref{calculation}, we report on the calculation details relevant for the process at hand. 

\section{The FDR method in a nutshell} \label{The_FDR_method}

FDR is a new method -gauge-invariant, free of infinities and 4-dimensional- to tackle loop calculations in QFTs. Here we review its main features and definitions, and refer to~\cite{Pittau:2012zd} for a more detailed discussion. 

Starting from the observation that UV divergences are unphysical and that they can be decoupled from the physically-relevant information of an amplitude, a new and coherent definition of loop integrals is possible, such that infinities are removed straight away and the integral is finite in 4-dimensions, with no need to absorb divergent quantities into renormalized parameters. 
Consider, for example, the dimensionally regulated integral 
\be
\label{eq:logdivint}
	\I =\int\dr \frac{1}{\D_0 \D_1}\,,
\ee
where $n=4+\ep$ is the number of space-time dimensions and
\bqa
\label{defd}
D_i ~=~ q^2-d_i\,,~~d_i ~=~ M^2_i-p^2_i- 2 (q\cdot p_i)\,,~~~{\rm with}~~~p_0 = 0\,.
\eqa
By introducing a new arbitrary  scale $\mu$, and defining 
\bqa
\Dbar_i~=~\D_i-\mu^2\,,~~\qbar^2~=~q^2-\mu^2\,,
\eqa
the logarithmically divergent part can be separated from the the physical one with the help of the following \emph{partial fraction identity}
\bqa
\label{partial_fractional_identities}
	\frac{1}{\Dbar_i}  = \frac{1}{\qbar^2}
		\Bigg(1+\frac{d_i}{\Dbar_i} \Bigg)	\,,
\eqa
yielding
\be
	\I = \lim_{\mu\rightarrow0}\;  \int \dr\, \frac{1}{\Dbar_0 \Dbar_1}\,
	= \lim_{\mu\rightarrow0}\; \int\dr \,\Bigg(\; 
			\Bigg[\frac{1}{\qbar^4}\Bigg]
			+\frac{d_1}{\qbar^4\Dbar_1}
			+\frac{d_0}{\qbar^2\Dbar_0\Dbar_1} 
			\;\Bigg)\,,
\ee
where the divergent piece is written between squared brackets. 
The $\mu$ scale is essential in order to regularize the integral in the low energy regime, and it is understood as the scale with respect to which loop momenta are very large: this is why infinities are expected to decouple in the limit $\mu \to 0$. Notice that the divergent contribution, being process-independent, effectively behaves like a vacuum bubble, and as such should not be taken into account when calculating a physical observable. The FDR integral corresponding to the integrand of eq.~(\ref{eq:logdivint}) is then {\em defined} to be
\bqa
\I^{\text{FDR}} = \int [ \de^4 q_i] \frac{1}{\Dbar_0 \Dbar_1} \equiv
	\lim_{\mu\rightarrow0}\; \int\de^4 q \,\Bigg(\; 
			 \frac{d_1}{\qbar^4\Dbar_1}
			+\frac{d_0}{\qbar^2\Dbar_0\Dbar_1} 
			\;\Bigg)\ \Bigg|_{\mu=\mu_R} ,
\eqa
so that it only includes the contribution conveying the physical information. 
The final identification $\mu= \mu_R$ effectively eliminates the dependence on the original cut-off, as explained in~\cite{Pittau:2012zd}.
This reasoning can be generalized to $\ell$ loops. Starting from a generic DR regularized integral
\be
	\I^{\text{DR}}_{\ell} = 
	\mu_R^{-\ell\ep}\int \prod_{i=1}^\ell\de^n q_i\;
	J(\{q^2\})\,,
\ee
where $\{q^2\}$ denotes the set of the squares of all possible combinations of loop-integration variables, the integrand $J = J_V + J_F$ can be parametrized in terms of $\mu$ and split into vacuum configurations ($J_V$) and finite part ($J_F$), so that one can write the FDR integral over $J$ as 
\bqa
       \label{eq_FDR}
	\I^{\text{FDR}}_{\ell} 
	&=&
	{\displaystyle \int \prod_{i=1}^\ell [ \de^4 q_i]}\;
	J(\{\qbar^2\}) 
	~\equiv~ \lim_{\mu\rightarrow0} \,
	{\displaystyle \int \prod_{i=1}^\ell\de^4 q_i}\;
	J_F( \{\qbar^2 \}) 
	\Bigg|_{\mu=\mu_R}
	, 
\eqa
where the replacement 
\be
\label{replacement}
\{q^2 \} \to \{\qbar^2\} 
\ee
is performed in both numerator and denominators.
The compact notation $\big[ \de^4q_i\big]$ in eq.~\eqref{eq_FDR} therefore implies:
\begin{enumerate}
\item parametrizing in terms of $\mu$;
\item separating and subsequently dropping the vacuum configurations;
\item integrating in 4 dimensions;
\item performing the limit $\mu\rightarrow 0$ until a logarithmic divergence is met;
\item evaluating the result in $\mu=\mu_R$.
\end{enumerate}
Note that one can also define the FDR integral as the difference between the DR integral and its vacuum configurations:
\be \label{FDR_as_a_difference}
	\I^{\text{FDR}}_{\ell} = \I^{\text{DR}}_{\ell} -
	\lim_{\mu\rightarrow0} \,
	\mu_R^{-\ell\ep}\int \prod_{i=1}^\ell\de^n q_i\;
	J_V (\{ \qbar^2\}) 
	\Bigg|_{\mu=\mu_R},
\ee
from which it is manifest the invariance under the shift of any integration variable.

Let us say a few more words on gauge-invariance, which is one of the key features of FDR, distinguishing it from other 4-dimensional methods. 
While it is obvious that terms conveying the kinematical dependence are equivalent in FDR and DR, a potential ambiguity remains in the constant term, because FDR subtracts vacuum configurations {\em before} integrating, while DR takes away poles in $\ep$ {\em after} the integration. In general the constant term can be fixed by enforcing gauge invariance as an extra constraint of the amplitude. However, FDR and DR alike automatically respect gauge invariance, thereby leading straightforwardly to the same correct constant. In order for gauge-invariance to be preserved in FDR, it is crucial that the parametrization in terms of $\mu$ is performed correctly, that is  propagators $\D_i$ and all $q^2$ appearing in the numerator should be promoted to their barred counterpart. This prescription is referred to as \emph{global treatment}. A $q$ and its associated $\mu$ should never be treated separately, which allows for the usual simplifications between numerator and denominator to take place. 
There are cases in which a $\mu^2$ does appear alone (we will discuss an example in the next section). In this type of terms, $\mu^2$ effectively plays the role of a squared loop momentum, and as such it should be treated. For example, the integral 
\be
\label{intmu}
\I^{\text{FDR}}(\mu^2) = 
	\int[\de^4q]\frac{\mu^2}{\Dbar_0\Dbar_1\Dbar_2}\,,
\ee
should be regarded as a logarithmically divergent one, that is to say 
it requires the {\em same }expansion needed to subtract the vacuum bubbles from
\bqa
\int[\de^4q]\frac{q^2}{\Dbar_0\Dbar_1\Dbar_2}\,,
\eqa
namely
\bqa
\label{exp3}
\frac{\mu^2}{\Dbar_0\Dbar_1\Dbar_2}= \mu^2\,\left(  
\left[\frac{1}{\qbar^6} \right]
+ \frac{d_0}{\qbar^2\Dbar_0\Dbar_1\Dbar_2}
+ \frac{d_1}{\qbar^4       \Dbar_1\Dbar_2}
+ \frac{d_2}{\qbar^6              \Dbar_2}\right)\,,
\eqa
giving
\be
\I^{\text{FDR}}(\mu^2) = 
   \lim_{\mu \to 0} \mu^2
     \left. \int d^4q 
               \left(
   \frac{d_0}{\qbar^2\Dbar_0\Dbar_1\Dbar_2}
  + \frac{d_1}{\qbar^4       \Dbar_1\Dbar_2}
 + \frac{d_2}{\qbar^6              \Dbar_2}
               \right) \right|_{\mu = \mur} \,.
\ee
The last integral behaves as $1/\mu^2$, so that, in the limit $\mu \to 0$ it does not vanish, but generates a finite constant:
\bqa
\I^{\text{FDR}}(\mu^2) = \frac{i \pi^2}{2}\,.
\eqa
This is equivalent to DR, when a finite term is obtained as the product of an $O(\ep)$-term and a single pole. More explicitly one can prove that 
\be
\label{correspondence}
	\int\dr \frac{(-\widetilde{q}^2)^k}{\D^{(n)}_0\D^{(n)}_1\ldots} 
	=
	\int [d^4q] \frac{(\mu^2)^k}{\Dbar_0\Dbar_1\ldots}\,,
\ee
where $\widetilde{q}^2 = (q^{(n)})^2-q^2$ is the $\ep$-dimensional part of an $n$-vector, and the superscript $(n)$ denotes an object living in $n$ dimensions.

 

As far as strings of Dirac matrices are concerned, the replacement rule $q^2\rightarrow \qbar^2$ is meant to be performed after calculating the trace (in 4 dimensions). The same result can be obtained by promoting $\slashed{q}\rightarrow \qbarslash \equiv q \pm \mu $ directly in the string \footnote{Thanks also to the fact that FDR integrals involving odd powers of $\mu$ vanish~\cite{Pittau:2012zd}.}.
This is achieved by defining $\qbarslash$ according to its position:
\be \label{eq_fermion_prescription}
	( \ldots 
		\, \qbarslash \; \gamma^{\alpha_1}\ldots\gamma^{\alpha_n}
		\qbarslash\, \ldots )
	= 
	(\dots 
		\, (\slashed{q}\pm\mu) \; \gamma^{\alpha_1}\ldots\gamma^{\alpha_n} (\slashed{q}\mp(-)^{n}\mu)  \ldots ) \,.
\ee
The sign within the first $\qbarslash$ is chosen arbitrarily; in the following $\qbarslash$, the sign is opposite if an even number of $\gamma$-matrices occur between the two $\qbarslash$'s, and it is the same in the case of an odd number of $\gamma$-matrices. Likewise should be treated the subsequent pairs of $\slashed{q}$'s occurring in the string. 
If chirality matrices are also involved, 
a gauge invariant treatment~\cite{Jegerlehner:2000dz} requires their anticommutation at the beginning (or the end) of open strings before replacing $\slashed{q}\rightarrow \qbarslash$. In the case of closed loops, $\gamma_5$ should be put next to the vertex corresponding to a potential non-conserved current. This reproduces the correct coefficient of the triangular anomaly, as observed in~\cite{Pittau:2012zd}. 


\section{Passarino-Veltman reduction in FDR} 
\label{PV_reduction_in_FDR}
The main asset of the Passarino-Veltman~\cite{Passarino:1978jh} method (PV) is that it provides a gauge-invariant decomposition of the amplitude: after the reduction, the full amplitude is expressed in terms of scalar integrals depending on physical thresholds only -multiplied by gauge independent coefficients- plus a rational part. 

In this section we review the PV reduction method within FDR. Our main aim is to illustrate the source of the constant terms contributing to the amplitude.
For simplicity, we explicitly work out the decomposition of a rank-2 bubble.
In the very same way higher rank tensors with more denominators can be decomposed, although, for brevity, we do not report here the result.

The integral
\be
	B^{\mu\nu} = \int \fdr \frac{q^{\mu}q^{\nu}}{\Dbar_0 \Dbar_1}\,,
\ee
can be decomposed in terms of  rank-2 symmetric tensors  depending on  $p \equiv p_1$:
\be
	B^{\mu\nu} = B_{00}\,g^{\mu\nu} + B_{11}\,p^{\mu}p^{\nu}\,.
\ee
By contracting both sides with $g_{\mu\nu}$ and $p_{\mu}p_{\nu}$, we obtain a system of equations from which the scalar coefficients $B_{00}$ and $B_{11}$ can be extracted:
\be \label{eq_PV_system}
	\left\{
	\begin{array}{r@{\;=\;}l}
		g_{\mu\nu}\; B^{\mu\nu} & 4 B_{00} + p^2 B_{11}\\
		p_{\mu}p_{\nu} \; B^{\mu\nu} & 
			p^2\,\big(B_{00} + p^2 B_{11} \big) \,.
	\end{array} \right.
\ee
With the usual algebraic manipulations we can express the left hand sides of eq.~\eqref{eq_PV_system} in terms of scalar integrals with 1 or 2 internal legs. Special care should be paid when contracting $B^{\mu\nu}$ with the metric tensor; indeed, 
\be
	g_{\mu\nu}\; B^{\mu\nu} = \int\fdr\frac{q^2}{\Dbar_0\Dbar_1}
	= \int\fdr \frac{\qbar^2+\mu^2}{\Dbar_0\Dbar_1}\,.
\ee
$\qbar^2$ takes part in the usual simplifications between numerator and denominator,
\be
	g_{\mu\nu}\; B^{\mu\nu} 
	= A_0(M_1^2)+M^2_0\,B_0(p^2; M_0^2,M_1^2) 
		+ \int\fdr \frac{\mu^2}{\Dbar_0\Dbar_1}\,,
\ee
while a $\mu^2$ is left out. 
The last integral should be treated as that in eq.~(\ref{intmu}) by expanding  $\frac{1}{\Dbar_0\Dbar_1}$ to extract all the terms behaving like $\frac{1}{\qbar^6}$. 
Alternatively, one can use the definition of the FDR integral as a difference 
of integrals, as in eq.~\eqref{FDR_as_a_difference} 
\footnote{The integration can be performed in 4 dimensions because the divergent integrands cancel in the difference.}:
\bqa
	\lim_{\mu\rightarrow 0}\,
        \int d^4q\, 
                     \Bigg\{
                     \frac{\mu^2}{\Dbar_0\Dbar_1}
              -\bigg[\frac{\mu^2}{\Dbar_0\Dbar_1}\bigg]_V
                     \Bigg\}\,,
\eqa
where 
\bqa
\bigg[\frac{\mu^2}{\Dbar_0\Dbar_1}\bigg]_V
= \mu^2\,\left(\frac{1}{\qbar^4}+\frac{d_0+d_1}{\qbar^6}+4\frac{(q\cdot p)^2}{\qbar^8}\right) \,,
\eqa
yielding
\bqa
	 \int\fdr\, \frac{\mu^2}{\Dbar_0\Dbar_1}
	= -\lim_{\mu\rightarrow 0}\,
        \int d^4q\, \mu^2
                   \Bigg\{
             \frac{d_0+d_1}{\qbar^6}+4\frac{(q\cdot p)^2}{\qbar^8}
                   \Bigg\}
	= \frac{i\pi^2}{2}\Big( M_0^2+ M_1^2-\frac{p^2}{3} \Big) \,.
\eqa
By solving the system in eq.~\eqref{eq_PV_system}, we finally obtain
\bqa
	B_{00} &=& -\frac{i\pi^2}{6}\Bigg(\frac{p^2}{3}-\Delta_+ \Bigg) \,
	+\, 
		\frac{A_0(M_0^2) \big[p^2-\Delta_-\big] 
			  + A_0(M_1^2) \big[p^2+\Delta_-\big]}{12\;p^2}\; \nl
			&&- \frac{B_0(p^2;M_0^2,M_1^2)
				\big[p^4-2\,p^2\,\Delta_+ +\Delta_-^2\big]}{12\;p^2}\,, 
				\nonumber
	\\
	B_{11} &=& \;\,\frac{i\pi^2}{6\,p^2}\Bigg(\frac{p^2}{3}-\Delta_+ \Bigg)
	\,+\, 
		\frac{-A_0(M_0^2) \big[p^2-\Delta_-\big] 
			  + A_0(M_1^2) \big[2p^2-\Delta_-\big]}{3\;p^4}\; \nl
 		&&+ \frac{B_0(p^2;M_0^2,M_1^2)
				\big[ (p^2-\Delta_-)^2-p^2\,M^2_0 \big]}{3\;p^4} \,,
\eqa
where
\be
	\Delta_{\pm} = M_1^2\pm M_0^2\,.
\ee
This result is consistent with that obtained with the standard PV reduction in DR.

Finally we point out that reduction methods other than PV can be used as well, e.g. the OPP method of~\cite{Ossola:2006us}. The basic observation is that any algebraic manipulation of the integrand is legal in FDR, the only subtlety being the replacement of eq.~(\ref{replacement}) in the numerator, which may generate a rational part \footnote{Called R2 in the OPP language.}. However, thanks to the correspondence in eq.~(\ref{correspondence}), this contribution can be reinserted back -in an OPP like reduction of the FDR integrals-  by using the same set of effective Feynman rules computed in DR~\cite{Draggiotis:2009yb,Garzelli:2009is}, or with the technique described in~\cite{Pittau:2011qp}. 

\section{The Higgs boson decay into two photons}
\label{calculation}
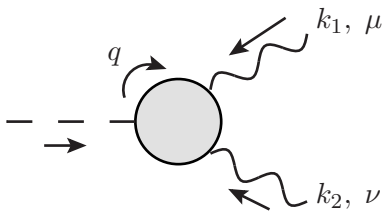
\begin{figure}
	\begin{center}
	\begin{picture}(176,88) (63,-57)
    \SetWidth{1.0}
    \SetColor{Black}
    \GOval(128,-23)(16,16)(0){0.882}
    \Line[dash,dashsize=10](64,-23)(112,-23)
    \Photon(140,-11)(176,10){4}{2}
    \Photon(140,-35)(176,-53){4}{2}
    \Text(180,10)[lb]{{\Black{$k_1,\;\mu$}}}
    \Text(180,-56)[lb]{{\Black{$k_2,\;\nu$}}}
    \Text(102,-2)[lb]{{\Black{$q$}}}
    \Arc[arrow,arrowpos=1,arrowlength=5,arrowwidth=2,arrowinset=0.2,clock](117,-11)(9.487,-161.565,-288.435)
    \Line[arrow,arrowpos=1,arrowlength=5,arrowwidth=2,arrowinset=0.2](162,-56)(150,-50)
    \Line[arrow,arrowpos=1,arrowlength=5,arrowwidth=2,arrowinset=0.2](168,16)(150,4)
    \Line[arrow,arrowpos=1,arrowlength=5,arrowwidth=2,arrowinset=0.2](78,-32)(90,-32)
  \end{picture}
	\end{center}
	\caption{\small Sketch of the Feynman diagrams contributing to the process; the loop can be either fermionic or bosonic (see figures~\ref{fermionic_diagrams} and~\ref{bosonic_diagrams}). The momenta are considered to be all incoming, and the virtual loop momentum $q$ is outgoing from the Higgs vertex. }
	\label{kinematics}
\end{figure}

Let us now concentrate on \hbox{$H \to \gamma\gamma$}. As mentioned above, because there is no $H\gamma\gamma$ vertex in the SM Lagrangian, this process is finite and loop-induced.
The generic amplitude is given by a bosonic and a fermionic contribution, which are independent of each other and separately gauge-invariant:
\be
	\amp(\beta,\eta) =  
		\Big( 
			{\amp}_W(\beta)
			+\sum_f N_c Q^2_f \,{\amp}_f(\eta)
		\Big)\,,
\ee
where $\beta$ and $\eta$ are dimensionless kinematic parameters defined as
\be
	\beta = \frac{4\,M_W^2}{M_H^2}\,, \qquad 
	\eta  = \frac{4\,m_f^2}{M_H^2}\,.
\ee
By denoting with $k_1$ and $k_2$ the momenta of the photons, as in figure~\ref{kinematics}, the amplitude reads
\be
	\amp = \amp^{\mu\nu}\,
		\varepsilon^*_{\mu}(k_1)\,\varepsilon^*_{\nu}(k_2)\, .
\ee
The tensorial structure of $\amp^{\mu\nu}$ is dictated by on-shellness and gauge invariance, i.e.
	\mbox{$k_i\cdot\varepsilon(k_i) = 0$} and 
	\mbox{$k_1^{\mu}\amp_{\mu\nu} = k_2^{\nu}\amp_{\mu\nu} = 0$},
so that
\be \label{eq_amp_form_factors}
	\amp^{\mu\nu}(\beta,\eta) = 	
	 	\Big( 
			\widetilde{\amp}_W(\beta)
			+\sum_f N_c Q^2_f \,\widetilde{\amp}_f(\eta)
		\Big)
		\;{T}^{\mu\nu}\;,
\ee
where $\widetilde{\amp}_W$ and $\widetilde{\amp}_f$ are scalar form factors of mass dimension -1, and 
\be
	{T}^{\mu\nu} = k_1^{\nu} k_2^{\mu} -(k_1\cdot k_2) \; g^{\mu\nu}\,.
\ee

\begin{figure}
\begin{center}
	\includegraphics[trim=0cm 7cm 0cm 7cm,clip=true,width=0.5\textwidth]{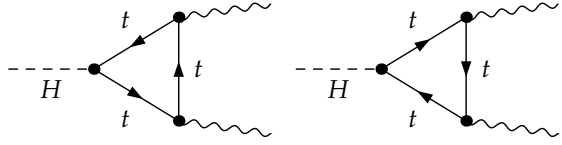}
	\caption{\small Diagrams contributing to the fermionic part of the amplitude (obtained with \FeynArts~\cite{Hahn:2000kx}).}
	\label{fermionic_diagrams}
\end{center}
\end{figure}

\begin{figure}
	\hspace{0.03cm}
	\includegraphics[trim = 0cm 3cm 0cm 3cm,clip=true,width=0.98\textwidth]{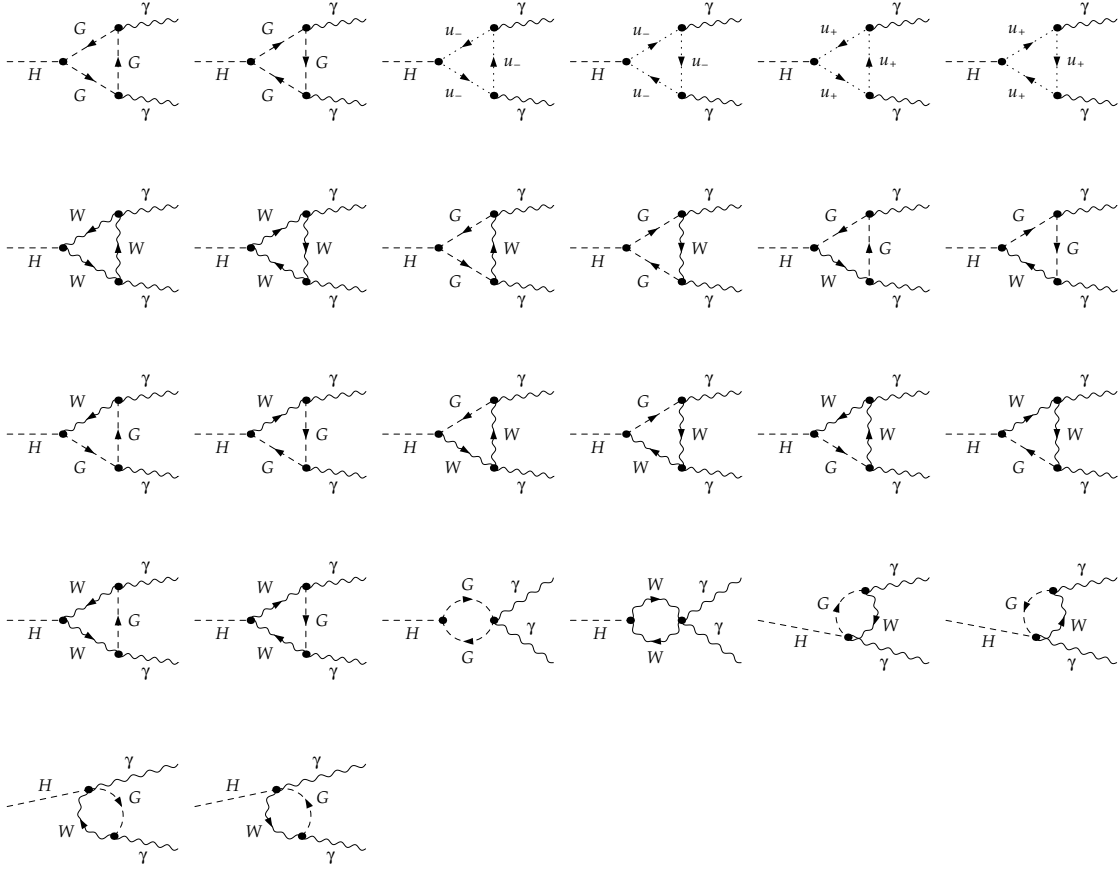}
\begin{center}
	\caption{\small Diagrams contributing to the bosonic part of the amplitude (obtained with \FeynArts~\cite{Hahn:2000kx}). $G$ denotes Goldstone bosons, while $u^{\pm}$ and $\overline{u}^{\pm}$ are the charged ghost and anti-ghost fields, respectively.}
	\label{bosonic_diagrams}
\end{center}
\end{figure}

$\widetilde{\amp}_W$ and $\widetilde{\amp}_f$ are obtained from the diagrams depicted in figures~\ref{fermionic_diagrams} and~\ref{bosonic_diagrams} with the Feynman rules of appendix~\ref{Feynman_rules}. Because we work in an arbitrary $R_\xi$ gauge, also ghost, scalar and mixed vectorial-scalar loops contribute to the bosonic part. Notice that diagrams can be distinguished according to the charge flow in the loop; however, because loops only couple to photons, graphs with the same topology but oppositely charged loops equally contribute to the amplitude. 
Schematically, naming the contribution of each diagram after the type of particles running in the loop, 
\bqa
\amp_f = \; 2\amp_{FFF}\,,
\eqa
as far as the fermionic contribution is concerned, and 
\bqa
	\amp_W &=& +2\amp_{SSS}+ 2\amp_{VVV}+ 2\amp_{SVS}+ 2\amp_{SSV}+
				2\amp_{VSS} \nl
	          &&+2\amp_{SVV}+ 2\amp_{VVS}+ 2\amp_{VSV}
			       +2\amp_{GGG_+}+ 2\amp_{GGG_-} \nl
		  &&+2\amp_{SV} + 2\amp_{VS} +\amp_{SS} +  \amp_{VV}\,,
\eqa
for the bosonic part. 
Because we work in the $R_{\xi}$-gauge, each bosonic diagram depends on the gauge parameter, e.g. \mbox{$\amp_{SV} = \amp_{SV}(\beta,\xi)$}, but their sum is gauge-invariant, i.e.  $\amp_W=\amp_W(\beta)$.

\begin{figure}
\begin{center}
\fcolorbox{white}{white}{
  \begin{picture}(101,71) (94,-43)
    \SetWidth{1.0}
    \SetColor{Black}
    \Line(95,-9)(128,-9)
    \Line(128,-9)(154,11)
    \Line(155,10)(155,-30)
    \Line(155,-29)(129,-9)
    \Line[dash,dashsize=5](155,10)(175,23)
    \Line[dash,dashsize=5](155,-29)(175,-42)
    \Text(160,-11)[lb]{{\Black{$m$}}}
    \Text(135,7)[lb]{{\Black{$m$}}}
    \Text(135,-30)[lb]{{\Black{$m$}}}
    \Text(107,-3)[lb]{{\Black{$m_0$}}}
  \end{picture}
}
\end{center}
	\caption{\small Scalar triangle $C_0(0,0,m_0;m,m,m)$; dashed lines refer to massless particles, solid lines to massive ones.}
	\label{triangle_topology}
\end{figure}
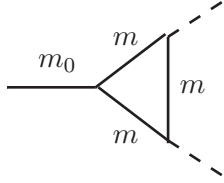

Even though the final result is finite, UV divergent loop integrals are encountered at intermediate steps. In particular, after simplifying reducible numerators, we deal with integrals with 2 or 3 internal legs and up to tensorial rank 2. Furthermore, the analytic structure of each diagram is characterized by a single threshold, either $M_W^2$ or $m_f^2$, or two thresholds $M_W^2$ and $\xi M_W^2$. 
To regularize the infinities, we have used the FDR method described 
in section~\ref{The_FDR_method}. Instead of calculating the integrals directly, we reduced the amplitude to scalar integrals by using the FDR version of PV reduction explained in section~\ref{PV_reduction_in_FDR}. The amplitude, expressed in terms of bubbles, self-energies and triangles ($A_0$, $B_0$ and $C_0$, respectively), is manifestly gauge-invariant, so that there remain only scalar integrals with a single physical threshold to be evaluated. In particular, all divergent integrals cancel out, and only scalar triangles must be computed, of the type given in figure~\ref{triangle_topology}.
Indeed, by combining the diagrams, performing the PV reduction, and evaluating the scalar integrals, we come by a result consistent with that  in~\cite{Marciano:2011gm}, which was obtained in DR, i.e.
\bqa
	\widetilde{\amp}_W(\beta) &=&
		\frac{i\,e^3}{(4\pi)^2s_W M_W}\;
		\Big[\, 
			2 + 3\beta + 3 \beta (2-\beta) f(\beta)
		\,\Big]\,,
	\label{eq_W_form_factor}
	\\
	\widetilde{\amp}_f(\eta) &=&
		\frac{-i\,e^3}{(4\pi)^2s_W M_W}\;
		2\eta\,\Big[\, 
			1 + (1-\eta) f(\eta)
		\,\Big]\,,
	\label{eq_f_form_factor}
\eqa
where $s_W=\sin\theta_W$ is the sine of the Weinberg mixing angle, and
\footnote{In eq.~(\ref{eqf}), $\varepsilon>0$ is a small imaginary part allowing for the analytic continuation of the result to any value of $x$.} 
\be
\label{eqf}
	f(x) = 
	-\frac{1}{4}\ln^2
		\Big( 
			\tfrac{1+\sqrt{1-x+i\varepsilon}}{-1+\sqrt{1-x+i\varepsilon}}
		\Big) 
		=
	\begin{cases}
		& 	\arctan^2\Big( \frac{1}{\sqrt{x-1}} \Big)
			\qquad\qquad\qquad \quad\,
			\text{if} \quad x\geq1\,
		\\
		&	-\frac{1}{4}\,
			\Big[ 
				\ln\Big(\frac{1+\sqrt{1-x}}{1-\sqrt{1-x}}\Big)
				-i\,\pi
			\Big]^2
			\;\;\,\,\,\qquad 
			\text{if} \quad x<1\,,
	\end{cases}
\ee
is a parametrization of the scalar triangle
\be
	C_0(0,0,s; M,M,M) = - \,\frac{2\,i\,\pi^2}{s}\, f(x)\,,
	\qquad x = \frac{4\,M^2}{s}\,.
\ee

Note that the bosonic form factor in eq.~\eqref{eq_W_form_factor} contains a constant term, independent of the kinematics. FDR automatically leads to its correct value, while in non-gauge-invariant frameworks it is necessary to enforce gauge-invariance~\cite{Piccinini:2011az,Dedes:2012hf} or momentum routing invariance~\cite{Cherchiglia:2012zp} as an extra constraint to recover it.


\subsection{The $W$ loop contribution}
 
\begin{figure}
\begin{center}
\fcolorbox{white}{white}{
  \begin{picture}(199,119) (95,-224)
    \SetWidth{0.5}
    \SetColor{Black}
    \Text(162,-139)[lb]{{\Black{$q$}}}
    \SetWidth{1.0}
    \Line[arrow,arrowpos=1,arrowlength=5,arrowwidth=2,arrowinset=0.2](216,-168)(216,-176)
    \Line[arrow,arrowpos=1,arrowlength=5,arrowwidth=2,arrowinset=0.2](185,-207)(169,-191)
    \Text(224,-176)[lb]{{\Black{$q+p_1$}}}
    \Text(139,-219)[lb]{{\Black{$q+p_2$}}}
    \Line[arrow,arrowpos=1,arrowlength=5,arrowwidth=2,arrowinset=0.2](170,-149)(186,-137)
    \Line[dash,dashsize=10](96,-172)(162,-172)
    \Line[dash,dashsize=10,arrow,arrowpos=0.5,arrowlength=5,arrowwidth=2,arrowinset=0.2,flip](162,-172)(205,-211)
    \Line[dash,dashsize=10,arrow,arrowpos=0.5,arrowlength=5,arrowwidth=2,arrowinset=0.2,flip](205,-211)(204,-145)
    \Photon(162,-172)(205,-141){5}{3}
    \Photon(205,-141)(256,-125){5}{3}
    \Photon(205,-211)(256,-219){5}{3}
  \end{picture}
}
\end{center}
	\caption{\small Vector-scalar-scalar loop diagram contributing to the bosonic part of the amplitude. }
	\label{VSS}
\end{figure}
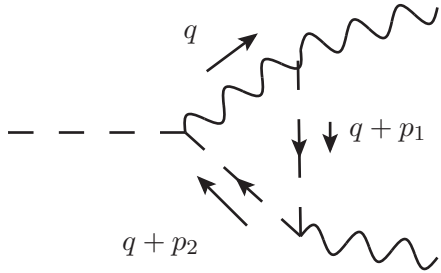

In this section, we work out in some detail the contribution to the amplitude coming from the diagram in figure~\ref{VSS}. This stands as an example to illustrate how to use the FDR method in practice. In particular, we are interested in showing that, thanks to the global treatment of the $\mu$ parameter, the integrand within an FDR integral can be algebraically manipulated just in the same way as any integrand of a standard 4- or $n$-dimensional loop integral. This fact,
together with the FDR shift invariance properties \footnote{See eq.~(\ref{FDR_as_a_difference}).}, preserves all cancellations required by gauge invariance.

The analytic contribution of the diagram in figure~\ref{VSS} to the amplitude $\amp_W$ is given by
\be
	\amp_{VSS}^{\mu\nu} = \int\fdr\, J_{VSS}^{\mu\nu}(\qbar^2)\,, 
\ee
where, according to the Feynman rules in appendix~\ref{Feynman_rules},
\be
		J_{VSS}^{\mu\nu}(\qbar^2)
		= \frac{e^3\,M_W}{2 \,s_W} 
		\frac{(2q+p_1+p_2)^{\nu}(q+2p_2)_{\rho}}
			{\Dbar_0\Dbar_1\Dbar_2}
		\bigg[
			g^{\mu\rho} - 
\frac{(1-\xi)\;q^{\mu}q^{\rho}}{(\qbar^2-\xi M_w^2)}
		\bigg]\,,
\ee
with $M^2_0 \equiv d_0= M_w^2$,  $M^2_1 = M^2_2 = \xi M_w^2$ and $p_n=\sum_{i=1}^n k_i$.
After contracting Lorentz indices, $q^2$ should be promoted to $\qbar^2$, as implied by the definition of the FDR integral.  The usual algebraic manipulations can then be performed on $J_{VSS}(\qbar^2)$. Consider, for example, the term
\be
\label{eq:a}
	\frac{\qbar^2}{\Dbar_0\Dbar_1\Dbar_2}\,,
\ee
which should equate
\be
\label{eq:b}
	\frac{1}{\Dbar_1\Dbar_2}
	+ \frac{d_0}{\Dbar_0\Dbar_1\Dbar_2} \,.
\ee
The integrands in eqs.~(\ref{eq:a}) and (\ref{eq:b}) give the same result upon
FDR integration if they generate the same remainder after removing the divergent vacuum configurations.
The three denominators in eq.~(\ref{eq:a}) can be rewritten as in eq.~(\ref{exp3}) 
\bqa
\label{eq:c}
	\frac{1}{\Dbar_0\Dbar_1\Dbar_2}
	= \Bigg[\frac{1}{\qbar^6}\Bigg]
		+ \frac{d_2}{\qbar^6 \Dbar_2} 
		+ \frac{d_1}{\qbar^4\Dbar_1\Dbar_2}
		+ \frac{d_0}{\qbar^2\Dbar_0\Dbar_1\Dbar_2}\,,
\eqa 
while only the first term in eq.~(\ref{eq:b}) needs an expansion (since the second one is finite):
\bqa
\label{eq:d}
	 \frac{1}{\Dbar_1\Dbar_2}  
	= \Bigg[\frac{1}{\qbar^4}\Bigg]
		+ \frac{d_2}{\qbar^4 \Dbar_2} 
		+ \frac{d_1}{\qbar^2\Dbar_1\Dbar_2}\,.
\eqa
The squared brackets indicate the vacuum configurations which need to be subtracted, so that the insertion of eqs.~(\ref{eq:c}) and (\ref{eq:d}) into  eqs.~(\ref{eq:a}) and (\ref{eq:b}), respectively, gives the expected result
\bqa
\label{eq:e}
	\int\fdr\, \frac{\qbar^2}{\Dbar_0\Dbar_1\Dbar_2}
 	&=& 
	\int\fdr\,  \Bigg(\frac{1}{\Dbar_1\Dbar_2} + 
				\frac{d_0}{\Dbar_0\Dbar_1\Dbar_2} \Bigg)
 \nonumber \\
	&=&  \lim_{\mu\rightarrow0}\; \int\de^4q\, 
	\,\Bigg\{ \, 
		  \frac{d_2}{\qbar^4 \Dbar_2} 
		+ \frac{d_1}{\qbar^2\Dbar_1\Dbar_2}
		+ \frac{d_0}{\Dbar_0\Dbar_1\Dbar_2}
	\,\Bigg\} \Bigg|_{\mu=\mu_R}\,, 
\eqa
and one can identify
\be
	\frac{\qbar^2}{\Dbar_0\Dbar_1\Dbar_2} = 
	\Bigg(\frac{1}{\Dbar_1\Dbar_2} + 
				\frac{d_0}{\Dbar_0\Dbar_1\Dbar_2} \Bigg)\,,
\ee 
at the integrand level.
It is important to realize that eq.~(\ref{eq:c}) reproduces the last integral in eq.~(\ref{eq:e})  {\em only if} the original $q^2$ appearing above the three denominators in eq.~(\ref{eq:a}) is also promoted to $\qbar^2$. Otherwise it would give
\bqa
	\lim_{\mu\rightarrow0}\; \int\de^4q\, 
	\,\Bigg\{ \, 
		  \frac{q^2 d_2}{\qbar^6 \Dbar_2} 
		+ \frac{q^2 d_1}{\qbar^4\Dbar_1\Dbar_2}
		+ \frac{q^2 d_0}{\qbar^2\Dbar_0\Dbar_1\Dbar_2}
	\,\Bigg\}  \Bigg|_{\mu=\mu_R}\,,
\eqa
which differs by a factor $i \pi^2/2$ from the correct result.

After simplifying all tensorial integrands appearing in $J_{VSS}$ to irreducible ones, we obtain
\bqa
	\amp_{VSS}^{\mu\nu}  
		&= \frac{e^3\,M_W}{2 \,s_W}
		&\,\Bigg\{\,
			\Big( 2\,C^{\mu\nu}_{2\,\text{thr}}
			+(p_1+p_2)^{\nu}\,C^{\mu}_{2\,\text{thr}}\, \Big)
			+ 2\,p_2^{\mu}\;
				\Big(2\,C^{\nu}_{2\,\text{thr}}\;
				+(p_1+p_2)^{\nu}\,C_{2\,\text{thr}} \;\Big)\,+ 
			\nonumber \\
			&& 
			- \bigg(\xi- \frac{p_2^2}{\Mw}\bigg)\;
			\bigg[\,
				2\,\Big(C^{\mu\nu}_{2\,\text{thr}}
				-C^{\mu\nu}_{1\,\text{thr}}\Big)
				+(p_1+p_2)^{\nu}\,
				\Big(C^{\mu}_{2\,\text{thr}}
				-C^{\mu}_{1\,\text{thr}}\Big)
			\bigg]+ 
			\nonumber\\
			&& 
			- \frac{1}{\Mw} 
			\bigg[ 
				2\,\Big(B^{\mu\nu}_{2\,\text{thr}}
				-B^{\mu\nu}_{1\,\text{thr}}\Big)
				+(p_1+p_2)^{\nu}\,
				\Big( B^{\mu}_{2\,\text{thr}}
				-B^{\mu}_{1\,\text{thr}} \Big)
			\bigg]
		\,\Bigg\}\,, \label{M_VSS_fin}
\eqa
where the remaining integrals 
\footnote{We put between parentheses the mass appearing in each propagator.} are scalar and tensorial bubbles and triangles in FDR, to be further reduced via PV reduction. In eq.~\eqref{M_VSS_fin} the subscript 
`$n$ thr' indicates the number of different thresholds; explicitly
\bqa
	C_{2\,\text{thr}}^{\mu_1\ldots \mu_r} &=& \int \fdr\, 
	\frac{q^{\mu_1}\ldots q^{\mu_r}}{\Dbar_0(\Mw)\Dbar_1(\xi\Mw)\Dbar_2(\xi\Mw)}\,,
	\nonumber \\
	C_{1\,\text{thr}}^{\mu_1\ldots \mu_r} &=& \int \fdr\, 
	\frac{q^{\mu_1}\ldots q^{\mu_r}}{\Dbar_0(\xi\Mw)\Dbar_1(\xi\Mw)\Dbar_2(\xi\Mw)}	
       \,,\nl
	B_{2\,\text{thr}}^{\mu_1\ldots \mu_r} &=& \int \fdr\, 
	\frac{q^{\mu_1}\ldots q^{\mu_r}}{\Dbar_0(\Mw)\Dbar_1(\xi\Mw)}\,,
	\nonumber \\
	B_{1\,\text{thr}}^{\mu_1\ldots \mu_r} &=& \int \fdr\, 
	\frac{q^{\mu_1}\ldots q^{\mu_r}}{\Dbar_0(\xi\Mw)\Dbar_1(\xi\Mw)}	
       \,.
\eqa

\subsection{The fermionic loop contribution}

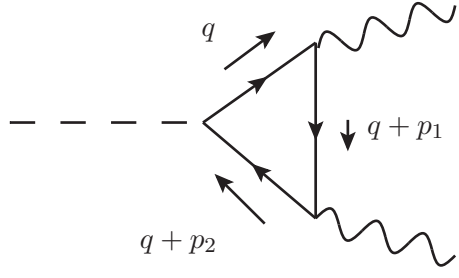
\begin{figure}
\begin{center}
\fcolorbox{white}{white}{
  \begin{picture}(205,119) (89,-224)
    \SetWidth{0.5}
    \SetColor{Black}
    \Text(162,-139)[lb]{{\Black{$q$}}}
    \SetWidth{1.0}
    \Line[arrow,arrowpos=1,arrowlength=5,arrowwidth=2,arrowinset=0.2](216,-168)(216,-176)
    \Line[arrow,arrowpos=1,arrowlength=5,arrowwidth=2,arrowinset=0.2](185,-207)(169,-191)
    \Text(224,-176)[lb]{{\Black{$q+p_1$}}}
    \Text(139,-219)[lb]{{\Black{$q+p_2$}}}
    \Line[arrow,arrowpos=1,arrowlength=5,arrowwidth=2,arrowinset=0.2](170,-149)(186,-137)
    \Line[dash,dashsize=10](90,-169)(156,-169)
    \Photon(205,-141)(256,-125){5}{3}
    \Photon(204,-205)(256,-219){5}{3}
    \Line[arrow,arrowpos=0.5,arrowlength=5,arrowwidth=2,arrowinset=0.2](162,-169)(204,-139)
    \Line[arrow,arrowpos=0.5,arrowlength=5,arrowwidth=2,arrowinset=0.2](204,-139)(204,-205)
    \Line[arrow,arrowpos=0.5,arrowlength=5,arrowwidth=2,arrowinset=0.2](204,-205)(162,-169)
  \end{picture}
}
\end{center}
	\caption{\small Fermionic loop diagram contributing to the amplitude. }
	\label{FFF}
\end{figure}

The contribution to the decay amplitude mediated by a loop of a charged fermion with mass $m_f$ is given by
\be
        \amp_{FFF}^{\mu\nu}\ 
	 = -\frac{e^3\,m_f}{2\,s_W\,M_W}
	\int [d^4q]\; 
	\frac{
	\text{Tr}\big[ (\qbarslash+\slashed{p}_2+m_f) \gamma^{\nu}
				  (\qbarslash+\slashed{p}_1+m_f) \gamma^{\mu}
				  (\qbarslash+m_f)
			 \big]
	}{\Dbar_0\,\Dbar_1\,\Dbar_2}\,,
\ee
with $M_0=M_1=M_2= m_f$, and where two types of traces containing twice the integration momentum appear, namely  
\be
	\begin{array}{rl} \label{eq_std_traces}
	\quad
	\Tr\big[\,\qbarslash\,\gamma^{\mu}\,\qbarslash\,\gamma^{\nu}\,\big]
	& = \Tr\big[\,(\slashed{q}\pm\mu)\,\gamma^{\mu}\,(\slashed{q}\pm\mu)\,\gamma^{\nu}\,\big]\,,
	\\
	\quad
	\Tr\big[\,\qbarslash\,\gamma^{\mu}\,\gamma^{\nu}\,\qbarslash\,\big]
	& = \Tr\big[\,(\slashed{q}\pm\mu)\,\gamma^{\mu}\,
			\gamma^{\nu}\,(\slashed{q}\mp\mu)\,\big]\,,
	\end{array}
\ee
with $\qbarslash$ defined in accordance with eq.~\eqref{eq_fermion_prescription}. 
This again allows all the usual algebraic manipulations at the level of the FDR
integrand; for example
\begin{eqnarray}
	\int [d^4q]\; 
	\frac{\Tr\big[\,\qbarslash\,\gamma^{\mu}\,\gamma^{\nu}\,\qbarslash\,\big]}{\Dbar_0\Dbar_1\Dbar_2}
	&=& 
	\int [d^4q]\; 
\frac{4 \,\qbar^2\,g^{\mu\nu}}{\Dbar_0\Dbar_1\Dbar_2}
\nonumber \\
        &=&
 4 \,g^{\mu\nu}
\left(
\int [d^4q]\frac{1}{\Dbar_1\Dbar_2}
+\int [d^4q]\frac{m_f^2}{\Dbar_0\Dbar_1\Dbar_2}
\right)\,.
\end{eqnarray}
After simplifying all reducible numerators we obtain
\begin{eqnarray}
	\amp_{FFF}^{\mu\nu}=-\frac{e^3\,m^2_f}{s_W\,M_W}
	\Big\{
		&-&2g^{\mu\nu}B + 
		\Big[ 2(p_1^{ \mu}p_2^{\nu}+p_2^{\mu}p_1^{\nu}) 
			-M_H^2g^{\mu\nu} 
		\Big] C  \nonumber \\
		&+& 4(p_1+p_2)^{\nu}C^{\mu} 
		+ 4 p_1^{\mu}C^{\nu}
		+ 8 C^{\mu\nu}\,
	\Big\}\,,
\end{eqnarray}
where
\be
C^{\mu_1\ldots\mu_r} =
\int\fdr\,
\frac{q^{\mu_1}\ldots q^{\mu_r}}{\Dbar_0\Dbar_1\Dbar_2}\,
~~{\rm and}~~
B =
\int\fdr\,
\frac{1}{\Dbar_0\Dbar_2}\,.
\ee
After a PV reduction of $\I^{\mu}_{3}$ and $\I^{\mu\nu}_{3}$ the result of eq.~\eqref{eq_f_form_factor} easily follows.


\section{Conclusions and outlooks}
\label{sec:concl}

We made use of the loop-mediated decay $H \to \gamma \gamma$ to illustrate the key features of FDR, the Four Dimensional and free of infinities Regularization/Renormalization approach recently introduced in~\cite{Pittau:2012zd}.
 In particular, we pointed out the mechanisms which lead to an automatic preservation of gauge invariance. To this aim, we performed the calculation in a generic $R_\xi$ gauge, showing that, unlike other four-dimensional methods, FDR naturally produces the correct rational part of the amplitude, with no need to impose extra constraints.

The same FDR mechanisms which respect gauge invariance at 1-loop are expected to work unchanged with more loops and in the presence of Infrared and/or Collinear divergences. The explicit verification of these assertions will be the subject of further investigations.
\appendix

\section{Feynman rules} \label{Feynman_rules}
We draw, in figure~\ref{frules}, the Feynman rules used throughout this work~\cite{Denner:1991kt}. The tensors $V_3^{\mu\nu\rho}$,  $V_4^{\mu\nu\rho\sigma}$ and the coupling constants are given by
\begin{align}
	V_3^{\mu\rho\sigma} 
	&
	= g^{\sigma\rho}(p_+-p_-)^{\mu}+ g^{\rho\mu}(p_+-q)^{\sigma} + g^{\mu\sigma}(q-p_-)^{\rho}\,;	
\\
\nonumber\\
	V_4^{\mu\nu\rho\sigma} 
	& 
	= 2g^{\mu\nu}g^{\sigma\rho}-g^{\sigma\mu}g^{\rho\nu}- g^{\sigma\nu}g^{\mu\rho}\,;
\\
\nonumber\\
	C_{SVV} 
	& = \left\{
				\begin{array}{l l}
					M_W/s_W 		&  	\quad \text{if} \quad 
									\text{$SVV$}=HW^+W^-\\
					-M_W 		&  	\quad \text{if} \quad
									\text{$SVV$}=G^{\pm}W^{\mp}\gamma
				\end{array} \right.;
\\
\nonumber\\
	C_{SSVV} 
	& = \left\{
				\begin{array}{l l}
					2	 		&  	\quad \text{if} \quad
									\text{$SSVV$}=\gamma\gamma G^+G^-\\
					-1/2s_W		&  	\quad \text{if} \quad
								\text{$SSVV$}=W^{\pm}\gamma G^{\mp}H
				\end{array} \right.;
\\
\nonumber\\
	C_{VS_1S_2} 
	& = \left\{
				\begin{array}{l l}
					-1	 		&  \quad\,\quad \text{if} \quad
									\text{$VS_1S_2$}=\gamma G^+G^-\\
					\mp \frac{1}{2 s_W}		
								&  \quad\,\quad \text{if} \quad
									\text{$VS_1S_2$}=W^{\pm} G^{\mp}H
				\end{array} \right..
\end{align}
$G$ denotes Goldstone bosons, while $u^{\pm}$ and $\overline{u}^{\pm}$ are the charged ghost and anti-ghost fields, respectively.

\begin{figure}[h!]
\begin{center}
\hspace{-1cm}
\fcolorbox{white}{white}{
\begin{picture}(396,390) (79,-39)
    \SetOffset(7,0)
    \SetWidth{0.5}
    \SetColor{Black}
    \SetWidth{1.0}
    
    \Line[arrow,arrowpos=0.5,arrowlength=5,arrowwidth=2,arrowinset=0.2](80,334)(128,334)
    \Text(145,326)[lb]{{\Black{$\FF$}}}
    
    \Line[dash,dashsize=7](224,334)(256,334)
    \Line[arrow,arrowpos=0.5,arrowlength=5,arrowwidth=2,arrowinset=0.2](256,333)(280,350)
    \Line[arrow,arrowpos=0.5,arrowlength=5,arrowwidth=2,arrowinset=0.2](280,318)(256,333)
    \Text(288,326)[lb]{{\Black{$\SFF$}}}
    
    \Photon(376,334)(408,334){3.5}{3}
    \Line[arrow,arrowpos=0.5,arrowlength=5,arrowwidth=2,arrowinset=0.2](408,333)(432,350)
    \Line[arrow,arrowpos=0.5,arrowlength=5,arrowwidth=2,arrowinset=0.2](432,318)(408,333)
    \Text(440,330)[lb]{{\Black{$\VFF$}}}
    
    \Photon(130,262)(178,262){3.5}{4}
    \Text(195,250)[lb]{{\Black{$\VV$}}}
        
    \Line[dash,dashsize=7](330,262)(378,262)
    \Text(394,254)[lb]{{\Black{$\ScSc$}}}    

    \Photon(130,190)(154,190){3.5}{2}
    \Photon(154,190)(178,174){3.5}{2}
    \Photon(154,190)(178,206){3.5}{2}
    \Text(195,187)[lb]{\small{\Black{$\VVVa$}}}
    \Text(130,182)[lb]{\footnotesize{\Black{$\mu$}}}
    \Text(178,166)[lb]{\footnotesize{\Black{$\rho$}}}
    \Text(154,198)[lb]{\small{\Black{$p_+$}}}
    \Text(154,174)[lb]{\small{\Black{$p_-$}}}
    \Text(138,198)[lb]{\small{\Black{$q$}}}
    \Text(178,210)[lb]{\footnotesize{\Black{$\sigma$}}}    

    \Photon(330,206)(354,190){3.5}{2}
    \Photon(330,174)(354,190){3.5}{2}
    \Photon(354,190)(378,206){3.5}{2}
    \Photon(354,190)(378,174){3.5}{2}
    \Text(378,208)[lb]{\footnotesize{\Black{$\sigma$}}}
    \Text(394,182)[lb]{\small{\Black{$\VVVV$}}}
    \Text(323,208)[lb]{\footnotesize{\Black{$\mu$}}}
    \Text(323,168)[lb]{\footnotesize{\Black{$\nu$}}}
    \Text(380,166)[lb]{\footnotesize{\Black{$\rho$}}} 
    
    \Line[dash,dashsize=7](130,118)(154,118)
    \Photon(154,118)(178,102){3.5}{2}
    \Photon(154,118)(178,134){3.5}{2}
    \Text(195,115)[lb]{\small{\Black{$\SVV$}}}
        
    \Photon(354,118)(378,134){3.5}{2}
    \Photon(354,118)(378,102){3.5}{2}
    \Line[dash,dashsize=7](330,134)(354,118)
    \Line[dash,dashsize=7](330,102)(354,118)
    \Text(394,115)[lb]{\small{\Black{$\SSVV$}}}
    
    \Line[dash,dashsize=7](154,46)(178,30)
    \Line[dash,dashsize=7](154,46)(178,62)
    \Text(181,62)[lb]{\small{\Black{$S_1$}}}
    \Text(181,25)[lb]{\small{\Black{$S_2$}}}
    \Photon(130,46)(154,46){3.5}{2}
    \Text(195,40)[lb]{\small{\Black{$\VSS$}}}
        
    \Line[dash,dashsize=7](352,46)(376,62)
    \Line[dash,dashsize=7](352,46)(376,30)
    \Line[dash,dashsize=7](330,46)(352,46)
    \Text(394,38)[lb]{\small{\Black{$\SSS$}}}

    \Line[dash,dashsize=2,arrow,arrowpos=0.5,
          arrowlength=5,arrowwidth=2,arrowinset=0.2](80,-26)(128,-26) 
    \Text(145,-34)[lb]{{\Black{$\GG$}}}   
    
    \Line[dash,dashsize=7](224,-26)(256,-26)
    \Line[dash,dashsize=2,arrow,arrowpos=0.5,arrowlength=4.167,
          arrowwidth=1.667,arrowinset=0.2](256,-26)(280,-10)
    \Line[dash,dashsize=2,arrow,arrowpos=0.5,arrowlength=4.167,
          arrowwidth=1.667,arrowinset=0.2,flip](256,-26)(280,-42)
    \Text(295,-34)[lb]{{\Black{$\SGG$}}}   
    \Text(283,-10)[lb]{\small{\Black{$\overline{u}^{\pm}$}}}
    \Text(283,-42)[lb]{\small{\Black{$u^{\pm}$}}}
        
    \Photon(368,-26)(408,-26){3.5}{3}
    \Line[dash,dashsize=2,arrow,arrowpos=0.5,arrowlength=4.167,
          arrowwidth=1.667,arrowinset=0.2](408,-26)(432,-10)
    \Line[dash,dashsize=2,arrow,arrowpos=0.5,arrowlength=4.167,
          arrowwidth=1.667,arrowinset=0.2,flip](408,-26)(432,-42)
    \Text(450,-30)[lb]{{\Black{$\VGG$}}}
    \Text(416,-15)[lb]{\small{\Black{$p$}}}
    \Text(437,-10)[lb]{\small{\Black{$\overline{u}^{\pm}$}}}
    \Text(437,-42)[lb]{\small{\Black{$u^{\pm}$}}}

\end{picture}
}
\end{center}
\vspace{.5cm}
\caption{\label{frules} Feynman rules relevant for computing $H \to \gamma \gamma$. External momenta are considered to be incoming.
} \label{F_Feynman_rules}
\end{figure}
 

\acknowledgments
This work was performed in the framework of the ERC grant 291377, ``LHCtheory: Theoretical predictions and analyses of LHC physics: advancing the precision frontier''. We also thank the support of the MICINN project FPA2011-22398 (LHC@NLO).

\bibliography{hpp}{}
\bibliographystyle{JHEP}
\end{document}